%% file: main.tex
\documentclass[5p,times,procedia]{elsarticle}

\usepackage{ecrc}
\usepackage[hidelinks]{hyperref}
\usepackage{footmisc}

\volume{00}

\firstpage{1}

\journalname{Procedia CIRP}

\runauth{J. Pfeiffer et al.}

\jid{trpro}

\AtBeginDocument{%
  \providecommand\BibTeX{{%
    \normalfont B\kern-0.5em{\scshape i\kern-0.25em b}\kern-0.8em\TeX}}}

\input{Imports.tex}

\begin{document}

\begin{frontmatter}
\input{P0.Title.tex}
\input{P1.Abstract.tex}
\end{frontmatter}


\input{01.Introduction.tex}
\input{02.Background.tex}

\input{03.RelatedWork.tex}
\input{05.Contribution.tex}

\input{06.CaseStudies.tex}
\input{07.Discussion.tex}
\input{08.Conclusion.tex}

\section*{Acknowledgements}
\label{sec:acknowledgement}
\vspace{-0.3cm}
\small
Partly funded by the Federal Ministry for Economic Affairs and Energy (BMWE) -- Factory-X -- 13MX001L. Website \url{https://factory-x.org}, the Federal Ministry for Economic Affairs and Energy (BMWE) -- growING -- 13IPC036G. Website \url{https://www.growingdigitaltwin.de}, and by the Innovation Campus Future Mobility - Website: \url{https://www.icm-bw.de/en/}.
%
\vspace{-1em}
\bibliography{main}
\bibliographystyle{elsarticle-harv}

\end{document}

%% file: Imports.tex
\input{Packages.tex}

\input{Acronyms.tex}

\input{Commands.tex}

\input{Edit.tex}

\input{Debug.tex}

\input{Listings.tex}

%% file: Packages.tex
\usepackage[utf8]{inputenc}
\usepackage{graphicx}     
\usepackage{amsmath}      
\usepackage{eurosym}
\usepackage{paralist}
\usepackage{mdframed}
\usepackage{blindtext}
\usepackage{soul}
\usepackage{comment}
\usepackage[TS1,T1]{fontenc}

\usepackage{enumitem}
\usepackage{enumitem}
\usepackage{multirow}
\usepackage[labelfont=bf]{caption}

\newenvironment{itemize*}%
  {\begin{itemize}%
    \setlength{\itemsep}{0pt}%
    \setlength{\parskip}{0pt}}%
  {\end{itemize}}
  
\usepackage{array}
\usepackage{longtable}

\newcolumntype{L}[1]{>{\raggedright\let\newline\\\arraybackslash\hspace{0pt}}m{#1}}
\newcolumntype{C}[1]{>{\centering\let\newline\\\arraybackslash\hspace{0pt}}m{#1}}
\newcolumntype{R}[1]{>{\raggedleft\let\newline\\\arraybackslash\hspace{0pt}}m{#1}}

\usepackage{enumitem}

\AtBeginDocument{%

}


%% file: Acronyms.tex
\usepackage[nomain,acronym]{glossaries}
\makeglossaries

\newcommand{\plural}[1]{\glspl{#1}\xspace}

\newcommand{\abkuerzung}[2]{\newacronym{#1}{#1}{#2}}

\abkuerzung{AADL}{Architecture Analysis \& Design Language}

\abkuerzung{ADL}{architecture description language}

\abkuerzung{ADR}{Action Design Research}

\abkuerzung{API}{application programming interface}

\abkuerzung{AS}{abstract syntax}

\abkuerzung{AST}{abstract syntax tree}

\abkuerzung{CBSE}{component-based software engineering}

\abkuerzung{CD}{Class Diagram}

\abkuerzung{CFG}{context-free grammar}

\abkuerzung{CoCo}{context condition}

\abkuerzung{CS}{concrete syntax}

\abkuerzung{CPS}{cyber-physical system}

\abkuerzung{DoE}{design of experiments}

\abkuerzung{DSL}{domain-specific language}

\newcommand{\dsls}{\plural{DSL}}

\abkuerzung{FSM}{finite-state machine}

\abkuerzung{GPL}{general-purpose programming language}

\abkuerzung{KR}{knowledge representation}

\abkuerzung{LCDP}{low-code development platform}

\abkuerzung{DPL}{DSL product line}

\abkuerzung{M2M}{model-to-model}

\abkuerzung{M2T}{model-to-text}

\abkuerzung{MBSE}{model-based systems engineering}

\abkuerzung{MDE}{model-driven engineering}

\abkuerzung{MDSE}{model-driven systems engineering}

\abkuerzung{OCL}{Object Constraint Language}

\abkuerzung{SC}{statechart}

\abkuerzung{Sem}{semantics}

\abkuerzung{SME}{small and medium-sized enterprise}

\abkuerzung{SLE}{software language engineering}

\abkuerzung{SOS}{Structural Operational Semantics}

\abkuerzung{SPL}{software product line}

\abkuerzung{SysML}{Systems Modeling Language}

\abkuerzung{TS}{technological space}

\abkuerzung{TSCA}{time-synchronous channel automata}

\abkuerzung{TSPA}{time-synchronous port automata}

\abkuerzung{UML}{Unified Modeling Language}

%% file: Commands.tex
\usepackage{xspace}
\usepackage{ifdraft}

\newcommand*{\ie}{\textit{i.e.,}\@\xspace}
\newcommand*{\eg}{\textit{e.g.,}\@\xspace}

\makeatletter
\newcommand*{\etc}{%
  \@ifnextchar{.}%
  {\textit{etc}}%
  {\textit{etc.}\@\xspace}%
}
\makeatother

\definecolor{se-green}{RGB}{0,128,0}
\definecolor{se-blue} {RGB}{0,0,204}

\newcommand{\code}[1]{\texttt{#1}}

\newcounter{papernumber}

\newcounter{requirement}[section]

\definecolor{MyBoxText}{RGB}{255,255,255}
\definecolor{MyBoxBG}{RGB}{13,50,153}

%% file: Edit.tex
\newboolean{edit}
\setboolean{edit}{false}

\ifthenelse{\boolean{edit}}{%

}{%
}

%% file: Debug.tex
\usepackage{ifdraft}
\usepackage{ifthen}

\newboolean{debug} 
\input{DebugSwitch.tex} 

\ifthenelse{\boolean{debug}}{
  \usepackage{showframe} 
  \usepackage{todonotes} 
  \newcommand{\xynote}[2]{\todo[inline]{#1: #2}}
  
  \newcommand{\smallnote}[2]{
    \par\noindent\makebox[\textwidth][c]{%
      \fbox{
        \begin{minipage}{0.9\textwidth}
          \scriptsize{\color{#1}{#2}}
        \end{minipage}
      }
    }
  }
  
  \newcommand{\hint}[1]{\smallnote{black}{#1}}
  \newcommand{\thought}[1]{\smallnote{teal}{#1}}
  \newcommand{\actionitem}[2]{\noindent$\circ$ #1 (#2)\\}
  \newcommand{\xactionitem}[2]{\noindent$\times$ #1 (#2)\\}
  
  \newcommand{\del}[1]{\textcolor{red}{\sout{#1}}} 
}{%
  \usepackage[disable]{todonotes}
  \newcommand{\xynote}[2]{}
  \newcommand{\smallnote}[1]{}
  
  \newcommand{\hint}[1]{}
  \newcommand{\thought}[1]{}
  \newcommand{\actionitem}[2]{}
  \newcommand{\xactionitem}[2]{}
  
  \newcommand{\del}[1]{} 
}

\DeclareUnicodeCharacter{301}{XXXXXXXXXXX}
\DeclareUnicodeCharacter{302}{XXXXXXXXXXX}
\DeclareUnicodeCharacter{308}{XXXXXXXXXXX}

%% file: DebugSwitch.tex
\setboolean{debug}{false}

%% file: Listings.tex
\usepackage{listings}
\newenvironment{ownfigure}[0]%
{\begin{figure}[htb!]}
{\end{figure}}

\usepackage{color}
\definecolor{DarkRed}{rgb}{0.75,0,0}
\definecolor{Lightgreen}{rgb}{0.588,1.0,0.588}
\definecolor{DarkGreen}{rgb}{0,0.5,0}
    
\lstdefinelanguage{MontiArc}[]{Java}{
  morekeywords={component, port, in, out, inv, package, import, connect, autoconnect}
}

\lstdefinelanguage{myJava}[]{Java}{
  commentstyle=\color{DarkGreen}\itshape 
}

\lstdefinelanguage{MontiArcAutomaton}[]{Java}{
  morekeywords={component, port, in, out, inv, package, import, connect,
  autoconnect, automaton, state, ocl, java, initial, final,
  noCompletion, chaosCompletion, var, mode, activate, transitions,
  modetransitions}, commentstyle=\color{DarkGreen}\itshape }

\lstdefinelanguage{MCConfig} { 
    morekeywords={config, Require, Model} 
}

\lstdefinelanguage{Manifest} { 
    morekeywords={Manifest, Bundle, ManifestVersion, Name, SymbolicName,
      Version, Require
    } 
}

\lstdefinelanguage{mcGrammar}[]{}{
  morekeywords={
    grammar, package, path, parser, lexer, nows, noslcomments, nomlcomments, 
    noident, nostring, noanything, nocharvocabulary, dotident, identrule,
    xmlcomments, hashcomments, texcomments, freemarkercomments, concept, 
    globalnaming, define, usage, options, true, false, protected, ident, 
    compilationunit
  }
}

\lstdefinelanguage{mcLng}[]{}{
  morekeywords={
    dsltool, language, package, path, parser, root, parsingworkflow, 
    rootfactory, lexer, nows, noslcomments, nomlcomments, noident, nostring,
    dotident, concept, globalnaming, define, usage, options, true, false, 
    protected, ident
  }
}

\lstdefinelanguage{mcManifest}[]{}{
  morekeywords={
    bundle, Bundle, Name, SymbolicName, true, false, Main, Class, 
    Version, Activator, Localization, Require, 
    Exclude, Eclipse, LazyStart, Vendor, Export, Package, 
    ClassPath
  }
}

\lstdefinelanguage{Alloy}[]{Java}{
commentstyle=\color{DarkGreen}\itshape,
  morekeywords={abstract,sig,->,fact,pred,fun,run,for,iff,
  not,no,one,all,some,lone,\#,set,in,and,or,but,exactly,none,univ,Int,assert,check},
  otherkeywords = {[2]????},
    morekeywords = {[2]????},
    keywordstyle={[2]\color{blue}},
    otherkeywords = {[3]????,<,<->,->, &, |, =, !=, !,<:,~},
    morekeywords = {[3]????,<,<->,->, &, |, =, !=, !,<:,~},
    keywordstyle={[3]\color{blue}}
  }

\lstdefinelanguage{mccd}[]{Java}{
  morekeywords={classdiagram,abstract,<<singleton>>,class,int,String,
  association,composition,extends}
}

\lstdefinelanguage{FreeMarker}[]{}{
  keywordsprefix={\#},
  keywords={in},
  commentstyle=\color{DarkGreen}\itshape }

\lstdefinelanguage{Mona}[]{}{
  morekeywords={ex0,all0,ex1,all1,ex2,all2,var0,var1,var2,pred,in,notin,include,union,inter,empty,assert},
  morecomment=[l]{\#},
  commentstyle=\color{DarkGreen}\itshape,
  otherkeywords = {[2]????,next,boolean,init,case,esac},
  morekeywords = {[2]????,next,boolean,init,case,esac},
  otherkeywords = {[3]????,<,<=>,=>, &, |, =, !=, !},
  morekeywords = {[3]????,<,<=>,=>, &, |, =, !=, !},
}

\lstdefinelanguage{myPython}[]{Python}{
  morekeywords={assert},
  morecomment=[l]{\#},
  commentstyle=\color{DarkGreen}\itshape,
}

\lstdefinelanguage{GeneratorConfiguration}[]{Java} {
  morekeywords={
    template, 
    generator, 
    ast, 
    runtime},
}

\lstdefinelanguage{ApplicationConfiguration}[]{Java} {
  morekeywords={
    application,
    behaviors,
    bindings,
    classdiagrams,
    components,
    factories,
    generators,
    map,
    to},
}

\lstdefinelanguage{Isabelle}[]{} {
    morekeywords={
        datatype,
        typedef},
}

%% file: P0.Title.tex
\dochead{35th CIRP Design Conference}%

\title{Declarative Policy Control for Data Spaces:\\ A DSL-Based Approach for Manufacturing-X}

\address[a]{Institute for Control Engineering of Machine Tools and Manufacturing Units, University of Stuttgart, Seidenstr. 36, 70174 Stuttgart, Germany}
\address[b]{FVA GmbH, Lyoner Str. 18, 60528 Frankfurt, Germany}
\address[c]{Robert Bosch GmbH, Robert-Bosch-Campus 1, 71272 Renningen, Germany}
\author[a]{Jérôme Pfeiffer\corref{cor}} 
\author[a]{Nicolai Maisch}
\author[b]{Sebastian Friedl}
\author[c]{Matthias Milan Strljic}
\author[a]{Armin Lechler}
\author[a]{Oliver Riedel}
\author[a]{Andreas Wortmann}


\aucores{*Corresponding author. Tel.: +49-711-685-84500; {\it E-mail address:} jerome.pfeiffer@isw.uni-stuttgart.de}

%% file: P1.Abstract.tex
\begin{abstract}
The growing adoption of federated data spaces, such as in the GAIA-X and the International Data Spaces (IDS) initiative, promises secure and sovereign data sharing across organizational boundaries in Industry 4.0. 
In manufacturing ecosystems, this enables use cases, such as cross-factory process optimization, predictive maintenance, and supplier integration. 
Frameworks and standards, such as the Asset Administration Shell (AAS), Eclipse Dataspace Connector (EDC), ID-Link and Open Platform Communications Unified Architecture (OPC UA) provide a strong foundation to realize this ecosystem. 
However, a major open challenge is the practical description and enforcement of context-dependent data usage policies using these base technologies—especially by domain experts without software engineering backgrounds. 
Thererfore, this article proposes a method for leveraging domain-specific languages (DSLs) to enable declarative, human-readable, and machine-executable policy definitions for sovereign data sharing via data space connectors. 
The DSL empowers domain experts to specify fine-grained data governance requirements—such as restricting access to data from specific production batches or enforcing automatic deletion after a defined retention period—without writing imperative code. 
\end{abstract}

\begin{keyword}
Data Spaces; Domain-Specific Languages (DSLs); Industry 4.0; Data Governance; Manufacturing-X




\end{keyword}

%% file: 01.Introduction.tex
\section{Introduction}
\label{sec:Introduction}

Modern manufacturing increasingly relies on cross-organizational collaboration. 
In networked production environments, stakeholders can concentrate on their core competencies while collectively improving quality, flexibility, and cost efficiency. 
To support such scenarios, the \textit{Manufacturing-X} initiative was launched to develop technology and conventions for trusted data ecosystems in which diverse stakeholders seamlessly exchange data across production networks \cite{PlattformI4.0.2022}.
Several approaches have demonstrated methods and software frameworks for enabling collaboration inside data spaces based on standards and conventions. 
However, the technological foundation for successful data space creation ultimately depends on interoperable interfaces that are straightforward to implement.
The \textit{Factory-X} project~\cite{factoryx_homepage} introduced the \textit{MX-Port} concept~\cite{mx_port_whitepaper}, an approach for interoperably combining different base technologies for data space connectors. 
The various MX-Port configurations presented so far rely on complex software stacks, creating barriers for domain experts who have limited programming expertise.
To address this challenge, we propose a base-technology-independent approach that focuses on the minimal, essential information required to implement MX-Ports in domain-specific applications. 
The contributions of this paper are: 
\begin{inparaenum}[(1)]
    \item An analysis of existing MX-Port implementation and their underlying technologies for the discovery and access \& usage layers, 
    \item a unifying metamodel that combines the information required for implementing the MX-Port in the technologies, ID Link, Asset Administration Shell (AAS), Open Platform Communications Unified Architecture (OPC UA), and Eclipse Data Space Connector (EDC), and 
    \item a prototypical implementation of this metamodel as a Domain-Specific Language (DSL) available in a GitHub repository~\cite{companion_repo} for declarative policy control of data spaces designed to enable the semi-automatic generation of interfaces for partners within a supply chain. 
\end{inparaenum}

The paper is structured as follows: \autoref{sec:Background} explains the general principles underlying this work. 
\autoref{sec:RelatedWork} outlines related research. 
The development of the proposed DSL is presented in \autoref{sec:Contribution} and demonstrated through a prototype in \autoref{sec:CaseStudy}. 
Finally, \autoref{sec:Discussion} discusses the results, and \autoref{sec:Conclusion} summarizes our findings and future outlook.

%% file: 02.Background.tex
\section{Background}
\label{sec:Background}
\subsection{Software Language Engineering}
Software language engineering~\cite{BPR+20} is a subfield of software engineering
that focuses on the design, implementation, and evolution of software languages. 
\dsls are a special form of software language that have a limited instruction set tailored to a specific application domain, such as digital twins~\cite{DHM+22}, internet of things~\cite{KRS+22}, security~\cite{HKR+22}, or mechanical engineering~\cite{BDH+20}. 
Engineering such domain-specific languages (DSLs)~\cite{BPR+20} entails defining three essential constituents~\cite{EGR12}, i.e., a context-free syntax, a collection of non-context-free validation procedures, and a definition of the language's behavior. 
In this paper we focus on the syntax definition of a DSL for specifying data space connectors. 
The syntax of a language can be defined in terms of metamodel~\cite{DBLP:conf/icsa/CombemaleBW17}  and grammars~\cite{BEK+19}. 
Metamodels, in most cases, specify the abstract syntax of a language. 
Similar to UML class diagrams they specify classes, attributes, and relationships between classes. 
In contrast to metamodels, grammars can be used to define both the concrete syntax of a language, which specifies how the language constructs are represented in text, as well as the abstract syntax, which specifies the structure of the language constructs. 
In the evaluation we will leverage an implementation of our unifying metamodel in a textual DSL based on a grammar-based syntax specification~\cite{companion_repo}. 

\subsection{Data Spaces}
A Data space is a trusted, standardized framework for secure, sovereign data sharing between organizations, letting companies exchange information without losing ownership or control~\cite{DBLP:conf/pods/HalevyFM06}.
It exists to unlock the value of cross-company data collaboration~\cite{DBLP:journals/electronicmarkets/MollerJSGSGGO24}, like improving supply chains or enabling predictive maintenance, while avoiding the risks of central platforms, lock-in, or loss of sensitive data.
It works by keeping data with its owner, linking participants through certified data space connectors that enforce machine-readable usage policies, contracts, and secure transfer, all under a shared governance and certification model. 
Recently, the initiative Factory-X~\cite{factoryx_homepage} plans to create an open and collaborative digital ecosystem for factory outfitters and operators. 
%
\begin{figure}[tb]
  \centering
  \includegraphics[width=\linewidth]{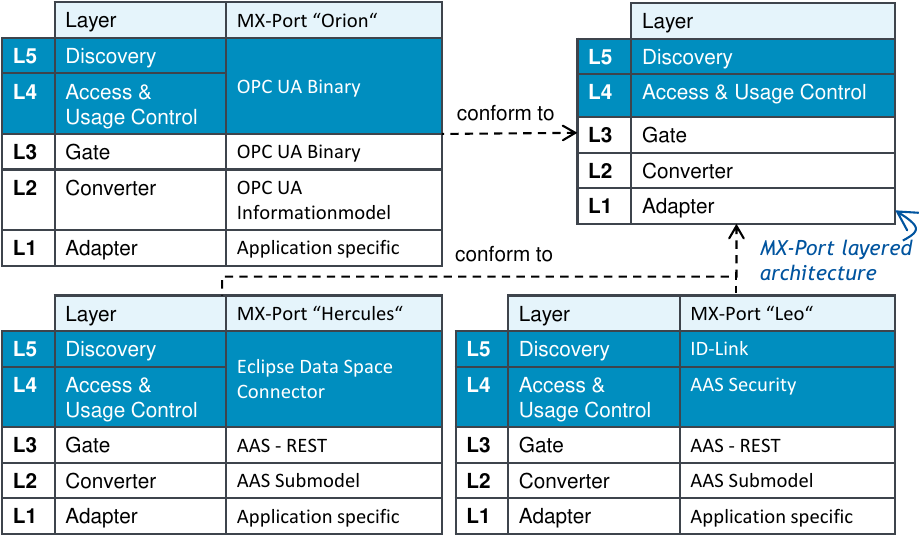}
  \caption{
    The five-layered MX-port architecture together with three application-specific implementations~\cite{mx_port_whitepaper}.
  }
  \label{fig:background_mx_port}
\end{figure}
The MX-Port concept, as defined by the Factory-X project~\cite{mx_port_whitepaper}, proposes a modular, five-layered architecture for federated data sharing across industrial ecosystems. 
The architecture (see \autoref{fig:background_mx_port}) includes five layers, where each layer represents a distinct functional responsibility in the data exchange process, and every MX-Port configuration is composed of components mapped to these layers: 
\begin{inparaenum}[(L1)]
    \item Adapter: Connects arbitrary business applications to the MX-Port, translating application-specific interfaces into standardized forms.
    \item Converter: Performs semantic alignment by mapping exchanged information to common models, such as Asset Administration Shell (AAS) submodels.
    \item Gate: Implements the uniform exchange interface for data transfer, ensuring compatibility and integrity between participants.
    \item Access \& Usage Control: Enforces permissions, licensing rules, and usage constraints defined by the data owner.
    \item Discovery: Provides mechanisms for finding partners, data assets, and relevant services within the ecosystem. 
\end{inparaenum}

%% file: 03.RelatedWork.tex
\section{Related Work}
\label{sec:RelatedWork}

Data spaces are designed to enable collaboration across company boundaries.  
Interfaces are therefore necessary to bridge these borders~\cite{NEUBAUER20231, Moreno.2023, DIN_RAMI.2016, Pivoto.2021}.
%
 A vision of a holistic Manufacturing-X architecture~\cite{NEUBAUER20231} suggests combining OPC UA, the AAS, and the EDC for vertical and horizontal integration. 
Similarly, strategies for implementing data spaces~\cite{Inigo.2022} emphasize the lack of standards for integrating these respective protocols properly. 
To tackle this, requirements from the Open Digital Rights Language (ODRL) can be leveraged~\cite{Der_Sylvestre_Sidibe.2024} to formalize privacy and access rules across the standards. 
Other contributions take a broader approach and cluster different protocols and serialization standards into meta-functionalities~\cite{Singh.2023} to form industrial data spaces, while \cite{Noardo.2024} propose data space building blocks by mapping base technologies to derived core functionalities. 
This clustering and generalization provides a useful overview, but a derivation of a minimal subset of standards or information relevant for MX-Port configurations is missing.
At the data space level, implementation technologies such as the EDC—which follows the International Data Spaces (IDS) standard~\cite{IDSA-DataspaceConnector-UsageControl-2025} have been analyzed. 
These studies point out issues with software documentation and compatibility~\cite{Dam.2023} and emphasize the complexity of such solutions~\cite{Bacco-2024}. 
In addition, \cite{Pampus.2022} highlight that a common understanding of key data space functionalities (such as usage policies and control enforcement) is still lacking, which hinders broader adoption.
In summary, prior research has developed concepts for integrating different industry standards vertically across production levels and horizontally across company boundaries. 
Approaches to cluster and map data space core functionalities have improved the conceptual understanding of requirements. 
However, a formalized way to configure and interface the different technologies and standards—especially for MX-Port configurations at the data space level—remains missing.

%% file: 05.Contribution.tex
\begin{figure}[tb]
  \centering
  \includegraphics[width=\linewidth]{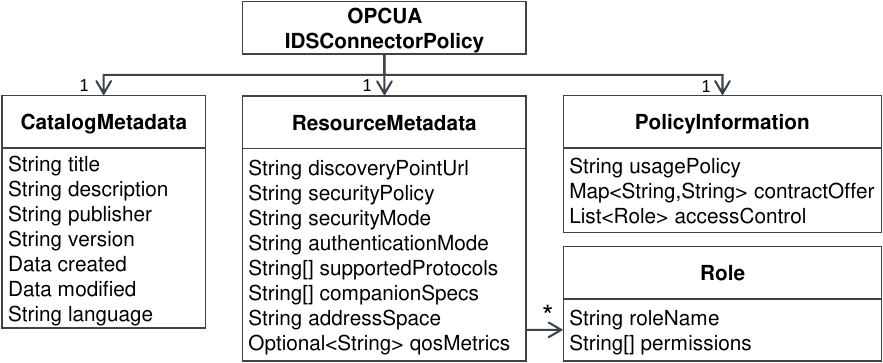}
  \caption{
    The metamodel of the information required to implement the layers L5 and L4 of MX-ports with OPC UA.
  }
  \label{fig:opc_ua_metamodel}
\end{figure}

\section{Towards a DSL for Declarative Policy Control for Data Space Connectors}
\label{sec:Contribution}
This section analyses the existing implementations of discovery and access \& control layers by OPC UA, EDC, and ID Link and the AAS. 
From this we derive a unified metamodel that can be leveraged to define a DSL for declarative policy control for data space connectors in these respective technologies. For the extraction of these metamodels we interviewed experts in the Factory-X research consortium and investigated implementations of the data space connectors from practice. 

\subsection{Analysing the OPC UA Metamodel}

OPC UA offers a toolset to implement the layers L4 and L5 of the MX-Port.
\autoref{fig:opc_ua_metamodel} shows the metamodel of the information required to realize these respective layers in an OPC UA data space connector implementation. 
For this, first, general \code{CatalogMetadata} is required. 
This entails defining a unique \code{title} of the offering, a \code{description} summarizing the the offering, the name of the \code{publisher}, \eg company XYZ GmbH, the \code{version} of the provided offering, \eg v1.0, 2025-Q3, and information such as when the offering was \code{created}, when it was last \code{modified}, and optionally a country code to specify in which language the offering is provided. 
For the specification of the layers L4 and L5, \code{ResourceMetadata} and \code{PolicyInformation} is required. 
The \code{ResourceMetadata} provides the URL of the OPC UA discovery endpoint, \eg \url{opc.tcp://machine-001.factory:4840}, as \code{dpointUrl}, the security policy, \eg Sha256, the security mode used for messaging, \eg Sign or SignAndEncrypt), and the authentication mode, \eg username or token-based.  
A list of supported communication protocols is also required, \eg OPC.TCP or MQTT. 
Furthermore, it needs to be specified which companion specifications are used via urls to the models, \eg machinery. 
In addition to that the address space defines the used OPC UA nodeset. 
Optionally, quality of service metrics, \eg sampling rate, maximum number of subscriptions, can be made explicit. 
A list of OPC UA roles specifies which role has the permission to access which kind of data in the data space. 
The \code{PolicyInformation} guides the access of roles via the attributes, \code{usagePolicy}, \code{contractOffer}, and \code{allowedRoles}. 
Usage policy refers to a IDS usage policy model~\cite{IDSA-DataspaceConnector-UsageControl-2025}. 
\code{contractOffer} defines a Map of key-value pairs that provide constraints such as \code{"validUntil": "2026-12-31"}. 
Finally \code{allowedRoles} associates roles with the policy.

\subsection{Analysing the Eclipse Data Space Connector Metamodel}
\begin{figure}[tb]
  \centering
  \includegraphics[width=0.65\linewidth]{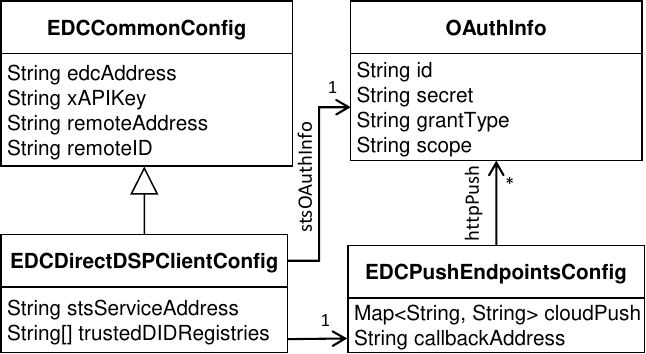}
  \caption{
    The metamodel of the information required to implement the layers L5 and L4 of MX-ports with Eclipse Data Space Connector.
  }
  \label{fig:edc_metamodel}
\end{figure}
For implementing the EDC the information depicted in \autoref{fig:edc_metamodel} is required for L4 and L5. 
An EDC can be hosted as a client. Then the EDC instance acts as an abstraction layer, that just needs to be configured with minimal connection details reflected in the metamodel class \code{EDCCommonConfig}. Then, EDC handles the Data Space Protocol (DSP) communication, so it is not necessary to directly manage low-level DSP authentication or token exchanges.
Besides, it is possible to connect directly to the DSP without the own hosted EDC. Then, it is necessary to implement and manage more of the protocol details that are contained in the class \code{EDCDSPClientConfig}. 
The \code{EDCCommonConfig} defines the address of the edc as an URL, a static password as the \code{xAPIKey}, the DSP address of the counter party as \code{remoteAddress} together with its \code{remoteID}, \ie the business partner number (BPN) or Decentralized identifiers (DID). 
For the advanced configuration the \code{EDCDirectDSPClientConfig} entails the definition of \code{stsServiceAddress}. STS (Security Token Service) refers to a service responsible for issuing security tokens that authenticate and authorize participants in a data space. 
Furthermore, the EDC with DSP can define a \code{EDCPushEndpointConfig} where clients can register and the data space connector can push data to.

\subsection{ID Link and Asset Administration Shell}
\begin{figure*}[ht]
  \centering
  \includegraphics[width=0.65\linewidth]{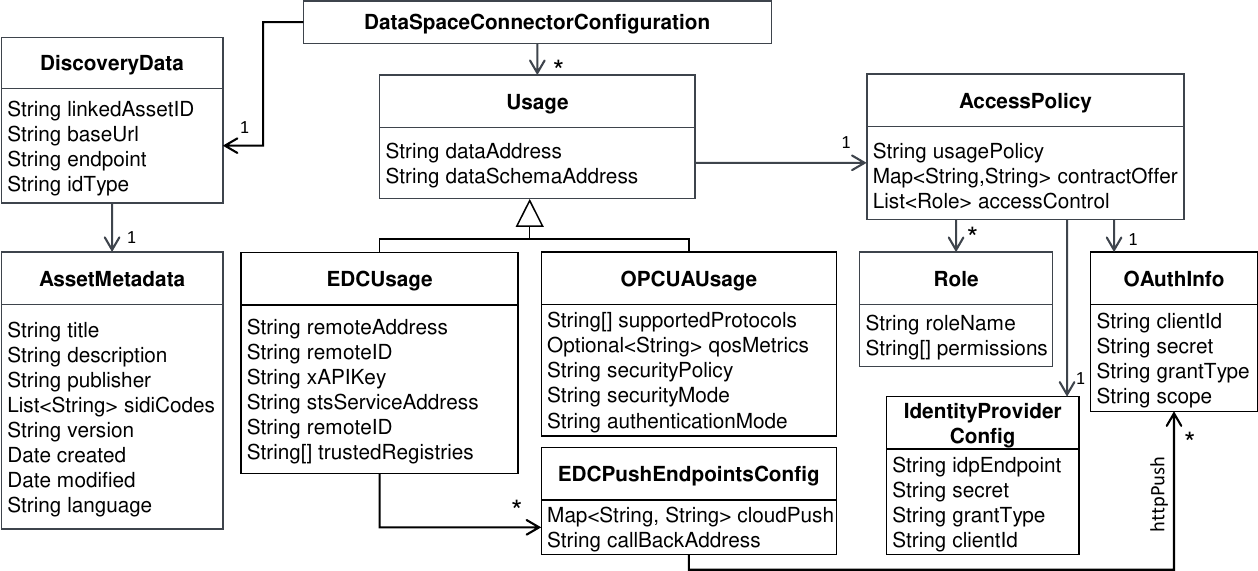}
  \caption{
    The unifying metamodel combining information required to implement the layers L5 and L4 of MX-ports with OPC UA, EDC, and ID Link \& AAS security.
  }
  \label{fig:unified_metamodel}
\end{figure*}

\begin{figure}[tb]
  \centering
  \includegraphics[width=0.95\linewidth]{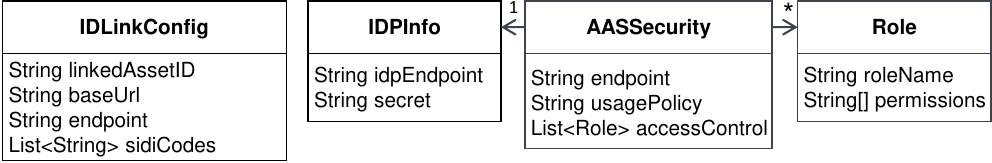}
  \caption{
    The metamodel of the information required to implement the layers L5 and L4 of MX-ports with ID Link and AAS security.
  }
  \label{fig:ID_Link_and_AAS_Security}
\end{figure}

For this configuration of a data space connector, ID link is used for the discovery layer, and AAS for the access \& usage layer. 
\autoref{fig:ID_Link_and_AAS_Security} depicts the information required for implementing both of these layers using the respective technology. 
The \code{IDLinkConfig} contains the globally unique id of the linked asset that should be discoverable over the \code{baseUrl} together with a unique \code{endpoint}. 
The system identificator (SIDI)~\cite{IEC61406-2} is the local identifier for the referenced asset, \ie serial number. 
The AAS security layer then controls who can access and use the data. 
For this, it requires an identity provider (IDP) with an endpoint and secret, and a usage policy that defines contracts and access control similar to OPC UA.

\subsection{Unifying Metamodel}

Based on the MX-Port architecture, the unifying metamodel defines the structure for describing the discovery, access and usage configuration data of data space connectors (Fig. \ref{fig:unified_metamodel}). 
The identification layer is represented by the class \code{IdentificationData}, which holds the essential identifiers of a resource such as the linked asset ID, base URL, endpoint, and identifier type. This information anchors the resource in the data space and connects it to its broader context. Building on this, the class \code{AssetMetaData} provides descriptive and administrative details, including title, description, publisher, semantic identifiers, versioning, creation and modification timestamps, and language. Together, these classes ensure that resources can be clearly described, catalogued, and discovered.

The usage layer is structured around the base class \code{Usage}, which defines the technical connection points to the resource, including the data address and schema address. 
This base class can be extended with technology-specific subclasses. 
\code{EDCUsage} specifies parameters relevant to EDC-based connectors, such as remote addresses, remote IDs, API keys, STS service addresses, and trusted DID registries, and can be complemented with \code{EDCPushEndpointsConfig} to define cloud push and callback configurations. 
\code{OPCUAUsage} models usage information for OPC UA resources, including supported protocols, quality-of-service metrics, security policies, message security modes, and authentication modes. These technology-specific extensions allow the metamodel to remain flexible and adaptable to heterogeneous technologies. 
Id-link is already covered by the base class properties \code{Usage}.  

The access and security layer is centered around the class \code{AccessPolicy}, which defines how resources may be consumed. 
This includes the usage policy, contractual clauses in the form of contract offers, and access control definitions. Roles capture a set of permissions tied to a role name, allowing fine-grained authorization. 
To support federated identity and secure integration with identity providers, the model includes \code{IdentityProviderConfig}, which defines identity provider endpoints, client credentials, grant types, and secrets, as well as \code{OAuthInfo}, which specifies parameters for OAuth-based authentication such as identifier, secret, grant type, and scope. 
Overall, the metamodel integrates descriptive metadata, discovery identifiers, technology-specific usage information, and access control mechanisms into a unified framework. It is extensible for different protocols and standards, while ensuring that discovery, usage, and access are consistently modeled and linked across the data space.

%% file: 06.CaseStudies.tex
\section{Case Study: Application to an Industrial Data Space}
\label{sec:CaseStudy}

Connected factories increasingly depend on heterogeneous participants that collaborate across digitally networked value chains. To investigate such interactions, a software-defined value network (SDVN) has been designed in which fictive companies represent key lifecycle phases, each realized through real machines and IT infrastructures at the University of Stuttgart~\cite{innowindow}.
The network includes a \textit{supplier} managing parts via robotic storage, a \textit{logistics provider} using a planar motion system, and \textit{manufacturers} offering complementary processes such as milling with robotic feeding or laser-based surface finishing. The \textit{assembly} is performed by collaborative robots, while the \textit{recycling} phase mirrors assembly by disassembling products and returning parts to supply.
This modular setting illustrates the heterogeneity of industrial participants and the need for reliable, sovereign data exchange across organizations. 
Here, data space connectors act as key enablers by bridging diverse infrastructures and protocols. 

\begin{figure}[tb]
  \centering
  \includegraphics[width=\linewidth]{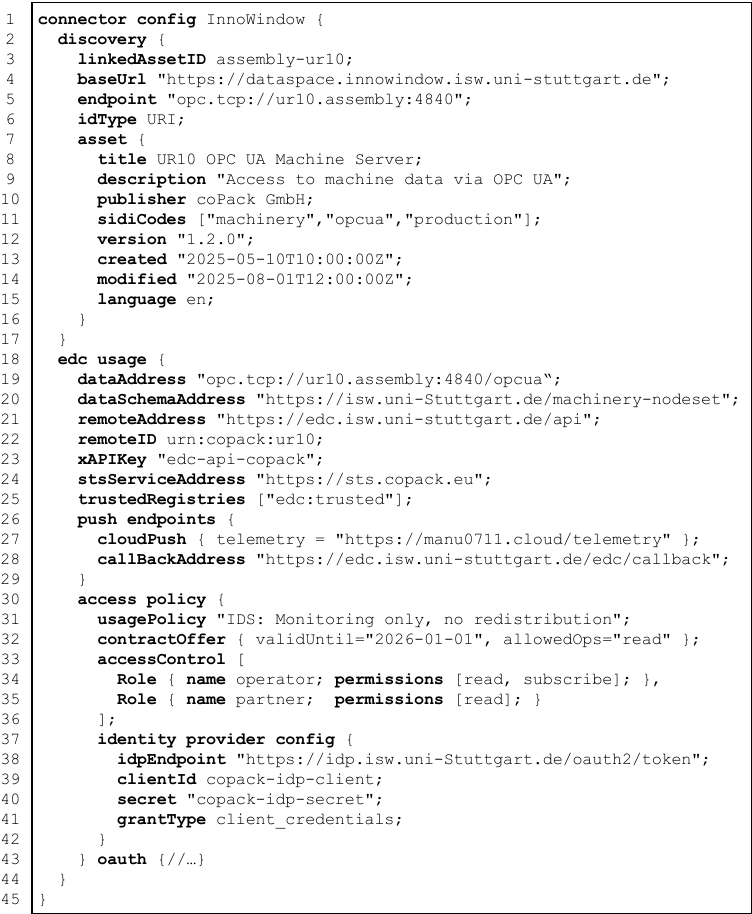}
  \caption{
    A model defining the EDC data space connector of a production machine. 
  }
  \label{fig:dsl_example}
\end{figure}

We implemented a prototype of our DSL for declarative data space connector descriptions and make it available online in a GitHub repository~\cite{companion_repo}. 
In the federated factory case study, our DSL demonstrates its value by providing a single, uniform way of describing connector configurations that would otherwise require fragmented and technology-specific setups. 
For space reasons we only depict one configuration (see \autoref{fig:dsl_example}) of one of the production line machines that is located at the manufacturer in the scenario described above. 
The production line uses a MX-Port configuration (see \autoref{fig:background_mx_port}), where the discovery and access \& usage layers are realized via EDC and the layers L3--L1 are handled by OPC UA. 
\autoref{fig:dsl_example} depicts the EDC configuration of L5 and L4 for a production line machine.
The model begins with the discovery data (ll.~2-17) that clearly links the connector to a specific machine asset in the factory. The asset metadata (ll.~7-16) then adds descriptive information such as the title, description, publisher, and version, making the resource both understandable for humans and discoverable by machines. 
On this basis, the section in lines 18-44 captures how the data can be used and accessed. 
In the case study, the usage layer is defined with EDC specifics, describing endpoints, API keys, and a push configuration. 
Access control (ll.~30-43) is modeled explicitly through policies, contracts, and roles. For example, the federated factory specifies that monitoring data is available only for trusted partners and that specific roles, such as "operator" or "partner," carry defined permissions (ll.~34-35). 
Finally, the model integrates security aspects by linking to an identity provider and specifying OAuth credentials (ll.~37-43). 
This ensures that the connector is not only technically reachable but also securely accessible within the federated environment.
The resulting specification highlights the practicality of the DSL. 
It enables engineers to describe all relevant aspects of connector configuration—discovery, usage, and access—in a single artifact that is both machine-processable and readable by humans. 

%% file: 07.Discussion.tex
\section{Discussion}
\label{sec:Discussion}

Our case study demonstrates the applicability of the DSL and the unified metamodel. However, several limitations remain. 
At this stage, the approach is purely analytical by comparing existing solution for data space connectors based on the MX-Port concept, and does not yet include a code generator that would translate models directly into executable configurations for connectors. 
Implementing this code generator is subject to future work within the Factory-X project. 
Another limitation lies in the scope of our analysis. The DSL so far is based on the technologies of OPC UA, EDC, and AAS with ID-Link. 
Extensions to other integration technologies, as well as to further architectural layers beyond discovery, usage, and access (e.g., monitoring, governance or intellectual property \cite{Annighoefer.2025} aspects), have not yet been realized. 
Furthermore, additional case studies would be necessary to assess the generalizability of the approach across different industrial settings and data space scenarios. 
Finally, the DSL does not yet support an import or modularization mechanism for reusing recurring configuration elements. 
Technically, however, this extension is possible through the language workbench MontiCore~\cite{DBLP:conf/sle/Pfeiffer023}, that we utilized for the implementation of the current DSL version. 
Besides, we plan to alleviate this limitation in the future by leveraging our expertise in language modularization and reuse~\cite{PW21} from previous publications in the field. 

%% file: 08.Conclusion.tex
\section{Conclusion}
\label{sec:Conclusion}
In this paper we analyzed three data space connector configurations according to the MX-Port layered-architectures and extracted the information that is necessary to implement the layers for discovery and access \& usage in OPC UA, AAS with ID-Link and EDC. 
From this analysis, we extracted metamodels for each technology and abstracted commonalities between those metamodels into a unifying metamodel, which can be extended with technology-specific aspects. 
We evaluated this metamodel by implementing a textual DSL and demonstrated its applicability in a SDVN. 
In the future, we plan to add a code generator to generate parts of connector implementations in the respective technologies and to extend our DSL to further layers of the MX-Port concept.

%% file: main.bib
@InProceedings{HKR+22,
    author= {Hacks, Simon and Katsikeas, Sotirios and Rencelj Ling, Engla and  iong, Wenjun and Pfeiffer, Jerome and Wortmann, Andreas},
    editor= {Augusto, Adriano and Gill, Asif and Bork, Dominik and Nurcan, Selmin and Reinhartz-Berger, Iris and Schmidt, Rainer},
    title= {Towards a Systematic Method for Developing Meta Attack Language Instances},
    booktitle= {Enterprise, Business-Process and Information Systems Modeling},
    year= {2022},
    publisher= {Springer},
    address= {Cham},
    pages= {139--154},
}

@inproceedings{BDH+20,
    key={BDH+20},
    author={Bibow, Pascal and Dalibor, Manuela and Hopmann, Christian and Mainz, Ben and Rumpe, Bernhard and Schmalzing, David and Schmitz, Mauritius and Wortmann, Andreas},
    editor={Dustdar, Schahram and Yu, Eric and Salinesi, Camille and Rieu, Dominique and Pant, Vik},
    title={{Model-Driven Development of a Digital Twin for Injection Molding}},
    booktitle={{International Conference on Advanced Information Systems Engineering}},
    location = {Grenoble},
    year={2020},
    month={June},
    publisher={Springer International Publishing},
    pages={85--100},
    series={Lecture Notes in Computer Science},
    volume={12127},
}

@inproceedings{BPR+20,
  author = {Butting, Arvid and Pfeiffer, Jerome and Rumpe, Bernhard and Wortmann, Andreas},
  title = {A Compositional Framework for Systematic Modeling Language Reuse},
  month = {October},
  year = {2020},
  publisher = {Association for Computing Machinery},
  address = {New York, NY, USA},
  pages = {35–46},
  numpages = {12},
  location = {Virtual Event, Canada},
  series = {MODELS '20},
  booktitle = {Proceedings of the 23rd ACM/IEEE International Conference on Model Driven Engineering Languages and Systems}
}

@article{DHM+22,
  title = {Generating customized low-code development platforms for digital twins},
  journal = {Journal of Computer Languages},
  volume = {70},
  pages = {101117},
  year = {2022},
  issn = {2590-1184},
  author = {Manuela Dalibor and Malte Heithoff and Judith Michael and Lukas Netz and Jérôme Pfeiffer and Bernhard Rumpe and Simon Varga and Andreas Wortmann},
}

@article{KRS+22,
    title = {{MontiThings: Model-driven development and deployment of reliable IoT applications}},
    journal = {Journal of Systems and Software},
    pages = {111087},
    year = {2022},
    issn = {0164-1212},
    author = {Jörg Christian Kirchhof and Bernhard Rumpe and David Schmalzing and Andreas Wortmann},
}

@article{BEK+19,
    key = {BEK+19},
    author = {Butting, Arvid and Eikermann, Robert and Kautz, Oliver and Rumpe, Bernhard and Wortmann, Andreas},
    journal={Journal of Systems and Software},
    volume = {152},
    pages = {50--69},
    title = {{Systematic Composition of Independent Language Features}},
    year = {2019},
    month = {Juni},
}

@inproceedings{PW21,
  title={Towards the Black-Box Aggregation of Language Components},
  author={Pfeiffer, J{\'e}r{\^o}me and Wortmann, Andreas},
  booktitle={2021 ACM/IEEE International Conference on Model Driven Engineering Languages and Systems Companion (MODELS-C)},
  pages={576--585},
  year={2021},
  organization={IEEE}
}

@inproceedings{DBLP:conf/sle/Pfeiffer023,
  author       = {J{\'{e}}r{\^{o}}me Pfeiffer and
                  Andreas Wortmann},
  editor       = {Jo{\~{a}}o Saraiva and
                  Thomas Degueule and
                  Elizabeth Scott},
  title        = {A Low-Code Platform for Systematic Component-Oriented Language Composition},
  booktitle    = {Proceedings of the 16th {ACM} {SIGPLAN} International Conference on
                  Software Language Engineering, {SLE} 2023, Cascais, Portugal, October
                  23-24, 2023},
  pages        = {208--213},
  publisher    = {{ACM}},
  year         = {2023}
}

@inproceedings{EGR12,
  title={{Language Composition Untangled}},
  author={Erdweg, Sebastian and Giarrusso, Paolo G and Rendel, Tillmann},
  booktitle={Proceedings of the Twelfth Workshop on Language Descriptions, Tools, and Applications},
  publisher = {ACM},
  doi = {10.1145/2427048.2427055},
  pages={1--8},
  year={2012},
  address = {Tallinn, Estonia}
}

@inproceedings{DBLP:conf/icsa/CombemaleBW17,
  author       = {Beno{\^{\i}}t Combemale and
                  Olivier Barais and
                  Andreas Wortmann},
  title        = {Language Engineering with the {GEMOC} Studio},
  booktitle    = {2017 {IEEE} International Conference on Software Architecture Workshops,
                  {ICSA} Workshops 2017, Gothenburg, Sweden, April 5-7, 2017},
  pages        = {189--191},
  publisher    = {{IEEE} Computer Society},
  year         = {2017}
}

@misc{PlattformI4.0.2022,
 editor = {{Federal Ministry for Economic Affairs and Climate Action}},
 author = {{Plattform Industrie 4.0}},
 year = {2022},
 title = {White Paper on Manufacturing-X},
 url = {https://www.plattform-i40.de/IP/Redaktion/EN/Downloads/Publikation/Manufacturing-X_long.html},
 address = {Berlin},
 urldate = {19.05.2025}
}

@article{NEUBAUER20231,
title = {{Architecture for manufacturing-X: Bringing asset administration shell, eclipse dataspace connector and OPC UA together}},
journal = {Manufacturing Letters},
volume = {37},
pages = {1-6},
year = {2023},
issn = {2213-8463},
author = {Michael Neubauer and Lukas Steinle and Colin Reiff and Samed Ajdinović and Lars Klingel and Armin Lechler and Alexander Verl}
}

@techreport{mx_port_whitepaper,
  title        = {MX-Port Concept V1.00},
  author       = {{Factory-X Project Team}},
  institution  = {Factory-X / Manufacturing-X Initiative},
  year         = {2025},
  month        = mar,
  day          = {27},
  url          = {https://factory-x.org/wp-content/uploads/MX-Port-Concept-V1.00.pdf},
  note         = {Accessed: 2025-08-14}
}

@article{innowindow,
 author = {Clar, Johannes and Dietrich, David and Nicolai Maisch and Reiff, Colin and Ziegler, Georg Heiner and Lechler, Armin and Verl, Alexander and Riedel, Oliver},
 title = {{Software-defined Value Network: An Industrial Testbed for Manufacturing-X (to appear)}},
 journal = {Proceedings of the 13th Congress of the German Academic Association for Production Technology (WGP), Hannover, September 2025},
 series = {Lecture notes in production engineering},
 series = {Lecture notes in production engineering},
 publisher = {Springer},
 year = {2025},
}

@article{DBLP:journals/electronicmarkets/MollerJSGSGGO24,
  author       = {Frederik M{\"{o}}ller and
                  Ilka Jussen and
                  Virginia Springer and
                  Anna Gie{\ss} and
                  Julia Christina Schweihoff and
                  Joshua Gelhaar and
                  Tobias Guggenberger and
                  Boris Otto},
  title        = {Industrial data ecosystems and data spaces},
  journal      = {Electronic Markets},
  volume       = {34},
  number       = {1},
  pages        = {41},
  year         = {2024}
}

@inproceedings{DBLP:conf/pods/HalevyFM06,
  author       = {Alon Y. Halevy and
                  Michael J. Franklin and
                  David Maier},
  editor       = {Stijn Vansummeren},
  title        = {Principles of dataspace systems},
  booktitle    = {Proceedings of the Twenty-Fifth {ACM} {SIGACT-SIGMOD-SIGART} Symposium
                  on Principles of Database Systems, June 26-28, 2006, Chicago, Illinois,
                  {USA}},
  pages        = {1--9},
  publisher    = {{ACM}},
  year         = {2006}
}

@misc{companion_repo,
   title = {Data Space Connector DSL},
  howpublished = {\url{https://github.com/jerome-pfeiffer1/FactoryX_DSL4DataSpaces}}, 
  year = 2025, 
  key = {GitHub Repository}
}

@misc{factoryx_homepage,
  author       = {{Factory-X Consortium}},
  title        = {Factory-X — An open and collaborative digital ecosystem for factory outfitters and operators},
  howpublished = {\url{https://factory-x.org/}},
  year         = {2025},
  note         = {Accessed: 2025-08-14}
}

@misc{IDSA-DataspaceConnector-UsageControl-2025,
  title        = {Usage Control},
  author       = {{International Data Spaces Association}},
  howpublished = {\url{https://international-data-spaces-association.github.io/DataspaceConnector/Documentation/v5/UsageControl}},
  year         = {2025},
  note      = {Accessed: 2025-08-14}
}

@misc{IEC61406-2,
  author       = {{International Electrotechnical Commission}},
  title        = {IEC 61406-2:2024 – Identification link – Part 2: Types/models, lots/batches, items and characteristics},
  year         = {2024},
  publisher    = {IEC},
  address      = {Geneva, Switzerland},
  month        = may,
  url          = {https://webstore.iec.ch/en/publication/77973},
  note         = {Accessed: 2025-08-15},
  isbn         = {9782832288863}
}

@ARTICLE{Moreno.2023,
  title     = {{Scalable Digital Twins for industry 4.0 digital services: a
               dataspaces approach}},
  author    = "Moreno, Tom{\'a}s and Almeida, Ant{\'o}nio and Toscano,
               C{\'e}sar and Ferreira, Filipe and Azevedo, Am{\'e}rico",
  journal   = "Prod. Manuf. Res.",
  publisher = "Informa UK Limited",
  volume    =  11,
  number    =  1,
  month     =  dec,
  year      =  2023,
  copyright = "http://creativecommons.org/licenses/by/4.0/",
  language  = "en"
}

@misc{DIN_RAMI.2016,
 key = {DIN SPEC 91345},
 year = {2016},
 title = {{Reference Architecture Model Industrie 4.0 (RAMI4.0)}},
 address = {Berlin},
 number = {DIN SPEC 91345},
 publisher = {Beuth}
}

@ARTICLE{Pivoto.2021,
  title     = {{Cyber-physical systems architectures for industrial internet of
               things applications in Industry 4.0: A literature review}},
  author    = "Pivoto, Diego G S and de Almeida, Luiz F F and da Rosa Righi,
               Rodrigo and Rodrigues, Joel J P C and Lugli, Alexandre Baratella
               and Alberti, Antonio M",
  journal   = "J. Manuf. Syst.",
  publisher = "Elsevier BV",
  volume    =  58,
  pages     = "176--192",
  month     =  jan,
  year      =  2021,
  language  = "en"
}

@INPROCEEDINGS{Singh.2023,
  title           = "Meta standard requirements for harmonizing dataspace
                     integration at the edge",
  booktitle       = "2023 {IEEE} Conference on Standards for Communications and
                     Networking ({CSCN})",
  author          = "Singh, Parwinder and {Nidhi} and Haq, Asim Ul and
                     Beliatis, Michail",
  publisher       = "IEEE",
  month           =  nov,
  year            =  2023,
  pages           = {130--135},
  conference      = "2023 IEEE Conference on Standards for Communications and
                     Networking (CSCN)",
  location        = "Munich, Germany"
}

@ARTICLE{Noardo.2024,
  title     = "Standards for data space building blocks",
  author    = "Noardo, Francesca and Atkinson, Rob and Bastin, Lucy and Maso,
               Joan and Simonis, Ingo and Villar, Alejandro and Voidrot,
               Marie-Fran{\c c}oise and Zaborowski, Piotr",
  journal   = "Remote Sens. (Basel)",
  publisher = "MDPI AG",
  volume    =  16,
  number    =  20,
  pages     = "3824",
  month     =  oct,
  year      =  2024,
  copyright = "https://creativecommons.org/licenses/by/4.0/",
  language  = "en"
}

@INPROCEEDINGS{Der_Sylvestre_Sidibe.2024,
  title           = "An approach for sovereign data exchange of {AAS} digital
                     twins through the international data space network",
  booktitle       = "2024 {IEEE} 29th International Conference on Emerging
                     Technologies and Factory Automation ({ETFA})",
  author          = "Sidibe, Gu{\'e}r{\'e}guin Der Sylvestre and Dhouib, Saadia",
  publisher       = "IEEE",
  volume          =  12,
  pages           = "1--4",
  month           =  sep,
  year            =  2024,
  conference      = "2024 IEEE 29th International Conference on Emerging
                     Technologies and Factory Automation (ETFA)",
  location        = "Padova, Italy"
}

@INPROCEEDINGS{Inigo.2022,
  title           = "Towards standardized manufacturing as a service through asset administration shell and international data spaces connectors",
  booktitle       = "48th Annual Conference of the {IEEE}
                     Industrial Electronics Society",
  author          = "Inigo, Miguel A and Legaristi, Jon and Larrinaga, Felix and Perez, Alain and Cuenca, Javier and Kremer, Blanca and
                     Montejo, Elena and Porto, Alain",
  publisher       = "IEEE",
  month           =  oct,
  year            =  2022,
  pages           = {1--6},
  conference      = "IECON 2022 - 48th Annual Conference of the IEEE Industrial
                     Electronics Society",
  location        = "Brussels, Belgium"
}

@inproceedings{Dam.2023,
  author       = {Tobias Dam and
                  Lukas Daniel Klausner and
                  Sebastian Neumaier and
                  Torsten Priebe},
  editor       = {Nicola Bena and
                  Beniamino Di Martino and
                  Antonio Maratea and
                  Alessandro Sperduti and
                  Emanuel Di Nardo and
                  Angelo Ciaramella and
                  Raffaele Montella and
                  Claudio A. Ardagna},
  title        = {A Survey of Dataspace Connector Implementations},
  booktitle    = {Proceedings of the 2nd Italian Conference on Big Data and Data Science
                  , Naples, Italy, September 11-13, 2023},
  series       = {{CEUR} Workshop Proceedings},
  volume       = {3606},
  publisher    = {CEUR-WS.org},
  year         = {2023},
}

@ARTICLE{Bacco-2024,
  title     = "What are data spaces? Systematic survey and future outlook",
  author    = "Bacco, Manlio and Kocian, Alexander and Chessa, Stefano and
               Crivello, Antonino and Barsocchi, Paolo",
  journal   = "Data Brief",
  publisher = "Elsevier BV",
  volume    =  57,
  number    =  110969,
  pages     = "110969",
  month     =  dec,
  year      =  2024,
  copyright = "http://creativecommons.org/licenses/by/4.0/",
  language  = "en"
}

@INCOLLECTION{Pampus.2022,
  title     = "Evolving data space technologies: Lessons learned from an {IDS}
               connector reference implementation",
  booktitle = "Lecture Notes in Computer Science",
  author    = "Pampus, Julia and Jahnke, Brian-Frederik and Quensel, Ronja",
  publisher = "Springer Nature Switzerland",
  pages     = "366--381",
  series    = "Lecture notes in computer science",
  year      =  2022,
  address   = "Cham"
}

@inproceedings{Annighoefer.2025,
  title     = {{Challenges of Collaborative MBSE in the Presence
of Practical Intellectual Property Requirements (to appear)}},
  author    = "Annighöfer, Björn and Maisch, Nicolai and Wortmann, Andreas",
  year = 2025,
  publisher = {{ACM}},
  booktitle    = {Proceedings of the {ACM/IEEE} 28th International Conference on Model Driven Engineering Languages and Systems, {MODELS} Companion 2024, Grand Rapids, USA, October 5-10, 2025}
}
